\documentstyle[prl,aps,multicol]{revtex}
\widetext

\renewcommand{\narrowtext}{\begin{multicols}{2} \global\columnwidth20.5pc}
\renewcommand{\widetext}{\end{multicols} \global\columnwidth42.5pc}
\def\Lrule{\vspace*{-0.2in}\noindent\vrule width3.5in height.2pt
  depth.2pt \vrule depth0em height1em}

\def\bml{\begin{mathletters}}
\def\eml{\end{mathletters}}
\def\beq{\begin{equation}}
\def\eeq{\end{equation}}
\def\bea{\begin{eqnarray}}
\def\eea{\end{eqnarray}}
\def\ba{\begin{array}}
\def\ea{\end{array}}

\def\to{\rightarrow}

\def\z{\zeta}

\begin{document}
\preprint{TIT-HEP-452}
\draft
\title{Comment on Dirac spectral sum rules for QCD$_3$}
\author{Shinsuke M. Nishigaki${}^*$}
\address{
Department of Physics, Faculty of Science,
Tokyo Institute of Technology,
Oh-okayama, Meguro, Tokyo 152-8551, Japan
}
\date{October 2, 2000}
\maketitle
\begin{abstract} 
Recently Magnea 
[Phys.\ Rev.\ {\bf D61}, 056005 (2000);
 Phys.\ Rev.\ {\bf D62}, 016005 (2000)]
claimed to have computed the first sum rules
for Dirac operators in 3D gauge theories
from 0D non-linear $\sigma$ models.
I point out that these computations are incorrect,
and that they contradict with the exact results
for the spectral densities unambiguously derived from
random matrix theory by Nagao and myself.
\end{abstract}
\narrowtext
Magnea \cite{Mag}
has recently claimed to have derived Dirac spectral sum rules for
three-dimensional gauge theories coupled
in a ($P, {\bf Z}_2$)-invariant manner to
fundamental fermions with $N_c=2$
(corresponding to the Dyson index $\beta=1$)
and adjoint fermions ($\beta=4$).
She employed
the small-mass expansion of the low-energy effective theories,
i.e. the zero-dimensional $\sigma$ models
over Riemannian symmetric spaces ${\cal M}=\,$CII
and BDI,
instead of AIII
that had been proposed
for the case with fundamental fermions and $N_c\geq 3$
($\beta=2$) \cite{VZ}.
She concluded that the first sum rule
in the presence of even number ($N_f$) of massless 2-component complex
or 4-component real (Majorana) fermions
is common both to the $\beta=1$ and $\beta=4$ universality classes,
and takes the form${}^\dagger$
\beq
\left\langle \sum_{i} \frac{1}{\z_i^2} \right\rangle^{(1,4)}=
\int_{-\infty}^\infty d\z \frac{\rho^{(1,4)}_s(\z)}{ \z^2}=\frac{4}{N_f},
\label{incorrect}
\eeq
where $\z$ stands for an unfolded Dirac eigenvalue
(i.e. rescaled by $1/(\pi {\rho}(0))$) and
$\rho^{(\beta)}_s(\z)$ stands for the scaled spectral density.
If true, this conclusion,
derived from an obviously correct formula (see Ref.\cite{SV})
\beq
\left\langle \sum_{i} \frac{1}{\z_i^2} \right\rangle
=\frac{d^2}{N_f \, M}
\label{obv}
\eeq
($d$ stands for the rank of the matrix that parameterizes ${\cal M}$,
and $M$ for the dimension of ${\cal M}$),
would be surprising,
as the four-dimensional
counterpart of the spectral sum rules are known to be
different for three values of $\beta$ \cite{LS,SV}.

On the other hand,
Nagao and myself \cite{NN}
have obtained Pfaffian expressions for the generic
$p$-level correlation functions
in a presence of an arbitrary number of (for $\beta=1$) and
an arbitrary number of pairwise degenerate (for $\beta=4$)
finite fermion masses $\{\mu_f\}$,
by applying the skew-orthogonal polynomial method
to pertinent random matrix ensembles.
To make comparison with Eq.(\ref{incorrect}),
I take a completely confluent limit $\mu_f\to 0$ for all $f$,
of our results 
[Ref.\cite{NN}, Eqs.(2.40), (2.42), (3.23), (3.25), (3.49), (3.51)]
with $p=1$ (spectral density), to obtain:
\widetext
\Lrule
\bml
\bea
\pi\,\rho^{(1)}_s(\z)
&=&
1 - \frac{3}{{\zeta }^2}+ \frac{3}{2 {\zeta }^4}  -
  \frac{3\,\cos 2 \zeta }{2 {\zeta }^4}
\ \ \ \ \ \ \ \ (N_f=2),\\
&=&
1 
- \frac{10}{{\zeta }^2}
- \frac{30}{{\zeta }^4}
+ \frac{210}{{\zeta }^6}
+ \frac{525}{2 {\zeta }^8}
+\left(
   \frac{70}{{\zeta }^5}
-   \frac{525}{{\zeta }^7}
\right)\sin 2 \zeta
+\left(
-   \frac{5}{{\zeta }^4}
 +   \frac{315}{{\zeta }^6}
-   \frac{525}{2 {\zeta }^8}
\right)\cos 2 \zeta
\ \ \ \ \ \ \ \ (N_f=4),
\\
&=&
1 
-   \frac{21}{{\zeta }^2}
- \frac{357}{2 {\zeta }^4}
-   \frac{945}{{\zeta }^6}
+ \frac{48195}{{\zeta }^8}
+ \frac{218295}{{\zeta }^{10}}
+ \frac{1964655}{2 {\zeta }^{12}}
+\left(
   \frac{378}{{\zeta }^5}
-   \frac{50652}{{\zeta }^7}
+   \frac{873180}{{\zeta }^9}
-   \frac{1964655}{{\zeta }^{11}}
\right)\sin 2 \zeta
\nonumber\\
&&
+\left(
-   \frac{21}{2 {\zeta }^4}
+ \frac{5859}{{\zeta }^6}
-   \frac{266490}{{\zeta }^8}
+   \frac{1746360}{{\zeta }^{10}}
-   \frac{1964655}{2 {\zeta }^{12}}
\right)\cos 2 \zeta
\ \ \ \ \ \ \ \ (N_f=6),
\eea
\eml
\bml
\bea
\pi\,\rho^{(4)}_s(\z)
&=&
1 - \frac{\sin 2 \zeta }{2 \zeta }
\ \ \ \ \ \ \ \ (N_f=2),\\
&=&
1 - \frac{{\sin^2 2 \zeta }}{4 {\zeta }^2}
+   
\left(
- \frac{\sin 2 \zeta}{4 {\zeta }^2}
+\frac{\cos 2 \zeta }{2 \zeta }
\right)
{\rm Si}\,(2 \zeta )
\ \ \ \ \ \ \ \ (N_f=4),\\
&=&
1 - \frac{3}{4 {\zeta }^2}+ \frac{3}{32 {\zeta }^4}
+   \left(
\frac{1}{{\zeta }}
-   \frac{3}{4 {\zeta }^3}
\right)
\sin 2 \zeta 
+   \frac{3\,\cos 2 \zeta }{2 {\zeta }^2}
-   \frac{3\,\cos 4 \zeta }{32 {\zeta }^4}
\ \ \ \ \ \ \ \ (N_f=6),
\\
&=&
1 
- \frac{27}{16 {\zeta }^2}
+ \frac{45}{64 {\zeta }^4}
+ \frac{45}{256 {\zeta }^6}
+\left(
\frac{15}{32 {\zeta }^3}
-   \frac{45}{64 {\zeta }^5}
\right)\sin 4 \zeta
+\left(
-   \frac{3}{16 {\zeta }^2}
+ \frac{45}{64 {\zeta }^4}
- \frac{45}{256 {\zeta }^6}
\right)\cos 4 \zeta
  \nonumber\\
&&  + 
\left[
\left(
\frac{9}{4 {\zeta }^2}
  - \frac{45}{32 {\zeta }^4}
\right)\sin 2 \zeta
+
\left( 
- \frac{3}{4 \zeta }
+  \frac{45}{16 {\zeta }^3}
  \right)\cos 2 \zeta
      \right]{\rm Si}\,(2 \zeta )
\ \ \ \ \ \ \ \ (N_f=8),
\eea
\eml
\narrowtext
\noindent
(Si stands for the sine-integral function) and so forth.
These expressions for the spectral densities
lead to the sum rules
\bea
&&\int_{-\infty}^\infty d\z \frac{\rho^{(1)}_s(\z)}{ \z^2}=
\frac{N_f}{(N_f-1)(2N_f+1)},
\label{correct1}\\
&&\int_{-\infty}^\infty d\z \frac{\rho^{(4)}_s(\z)}{ \z^2}=
\frac{N_f}{(N_f-1)(N_f/2+1)},
\label{correct4}
\eea
which are sensitive to the Dyson index $\beta$.
The above sum rules agree perfectly with
the numerical results for random matrix ensembles
of large but finite ranks $(\sim 40)$,
obtained by Hilmoine and Niclasen \cite{HN}
via two alternative methods
(an analytical method of Widom's [Table 4 of Ref.\cite{HN}] and
numerical Monte-Carlo simulations of random matrix ensembles).
Therefore I conclude that the expression (\ref{incorrect})
is erroneous,
and the coincidence of the first sum rules
for $\beta=1$ and $\beta=4$
claimed in her papers is illusory.\\

I thank C. Hilmoine for bringing this issue
to my attention.
This work was supported in part by JSPS, and
by Grant-in-Aid No.\ 411044
from the Ministry of Education, Science and Culture, Japan.

\widetext
\end{document}